%% file: lnicst.tex
\documentclass[lnicst]{svmultln}
\usepackage{makeidx}  
\usepackage{graphics}
\usepackage{booktabs}
\usepackage{todonotes}
\usepackage{url}

\begin{document}
\mainmatter              
\title{Mood-based On-Car Music Recommendations}
\titlerunning{Mood-based On-Car Music Recommendations}  
%
\author{Erion \c{C}ano\inst{1} \and Riccardo Coppola\inst{1} \and Eleonora Gargiulo\inst{2} 
\and Marco Marengo\inst{2} \and Maurizio Morisio\inst{1}}
\authorrunning{Erion \c{C}ano et al.}   
%

\institute{Department of Control and Computer Engineering , Polytechnic University of Turin, \\
Corso Duca degli Abruzzi, 24, 10129 Torino, Italy \\
\email{erion.cano@polito.it}
\and 
Joint Open Lab MobiLAB, Torino, Italy
}

\maketitle              

\begin{abstract}        %
Driving and music listening are two inseparable everyday activities for millions of 
people today in the world. Considering the high correlation between music, mood 
and driving comfort and safety, it makes sense to use appropriate and intelligent 
music recommendations based on the mood of drivers and songs in the context of 
car driving. 
The objective of this paper is to present the project of a 
contextual mood-based music recommender system capable of regulating the driver's mood 
and trying to have a positive influence on her driving behaviour. 
Here we present the proof of concept of the system and describe the techniques and technologies that are part of it. Further 
possible future improvements on each of the building blocks are also presented. 
\end{abstract}

\input{sections/Introduction}
\input{sections/Background}
\input{sections/SystemDescription}
\section{Discussion}
A great 
amount of the music streamed from the Internet today is suggested by 
intelligent recommender systems embedded in popular music portals such as 
TIMmusic, last.fm or Spotify. We have considered using a music 
recommender system also in a car environment, using the driver's mood and other contextual 
data to recommend musical tracks able to adjust the driver's mood for a 
comfortable and safe drive. 
In this work we presented the proof of concept of the system and its building 
blocks. The work is still in progress and we expect to have improvements or module 
extensions in the near future. We are working to adopt a finer grained mood 
representation model with more categories. Also the driver's mood recognition can 
be based on more than just heart rate dynamics (for instance, facial expressions and skin 
conductance can be additional alternatives) and can take into account external factors, like 
the speed of the vehicle compared to the speed limits of the road percurred (with a higher speed taken as a consequence of an excited mood, 
and a lower speed as a possible proof of drowsiness). Moreover, we are experimenting with 
multi-modal music mood classifiers which consider audio and textual feature of 
songs for better predictive and recommendation accuracy.  
\section{Acknowledgments}
This work was done at the Joint Open Lab MobiLAB and was supported by a fellowship 
from TIM.
%
%
%
\bibliographystyle{abbrv}
\bibliography{lnicst}  

\end{document}

%% file: sections/Introduction.tex
\section{Introduction}
Since the advent of the first car radios, listening to music has always been one of the 
favorite activities carried out by people while driving their cars: as reported 
in \cite{dibben2007exploratory}, about 70\% of the drivers do so. 
Driven by their tastes, attitudes and moods and by the nature of the trips, 
people have always selected the most adequate songs from their libraries 
creating their own customized playlists.
In \cite{schafer2013psychological} the authors attempt to answer the question 
``Why do people listen to music?''. Their finding is that the three most common 
motivations for listening to music are \emph{Self awareness}, 
\emph{Social relatedness} and \emph{Arousal and mood regulation}. 
These findings reveal a strong correlation between listening to music and self 
regulation of mood. 
Thus, people may want to confirm the mood state they are in by listening to music in 
accordance with it; conversely, they may want to get out of a bad mood by listening 
to songs capable of encouraging opposite feelings.
In the automotive domain there are many psychological studies that reveal 
connections between the played music and the actual behavior, concentration or 
performance of people driving their cars. 
Obviously, an appropriate background music stimulus may be an useful instrument to enhance 
the human performance and comfort in driving \cite{Nesbit, valenza2014revealing}. 
\par
On the other hand, the evolution of music recommender systems in the last decade 
has encouraged a high percentage of music listeners to rely on suggestions given 
by such applications. Music mood recommendations are embedded in many music 
social communities such as \emph{Last.fm}, \emph{TIMmusic} or \emph{Spotify}. 
This article presents the proof of concept of a Mood-Based On Car Music Recommender System
that uses various sources of information for tuning the recommendations: 
physiological information about the heart rate dynamics of the driver 
obtained from wearable sensors; user's saved musical preferences to take 
into account his/her musical tastes; telemetry of the current drive to consider the actual 
driving style of the user; location and time information to adapt the chosen playlists to 
the driving context. Using a tag-based folksonomy we classify the 
musical tracks in four categories: \emph{Happy}, \emph{Tender}, \emph{Sad}, \emph{Angry}.
Four sets of tracks are fed inside the recommendation module together with the other data and
the most convenient playlist is generated and recommended to the user. 
The rest of the paper is organized as follows: section 2 provides the necessary background 
about the principal building blocks of the project; 
section 3 illustrates the design of our project, with the choices we have made for each 
building block and the motivation behind them; finally, section 4 gives some hints for future 
extensions of the project. 

%% file: sections/Background.tex
\section{Background}
In this section we give insights about the state of the art of the components of 
this project. First we present some  models for mood representation, both for people 
and songs. We continue providing some relations between mood, music and driving 
safety. Then we describe what recommender systems are, and identify what contextual data 
can be utilized by them and how. Finally we give some information about in-car music 
infotainment systems, where our project of mood-based music recommendations 
will eventually be deployed.
\subsection{Modeling Emotions}
There are various definitions of mood in the works of psychiatrists that have 
relevance to this project. For example, 
while trying to point out the differences between \emph{Affect}, \emph{Emotion} 
and \emph{Mood} (which seem to be highly interrelated), the author 
in \cite{tenenbaummeasurement} provides the following: 
\begin{itemize}
	\item \textbf{Affect} - a neurophysiology state consciously accessible as a simple 
	primitive non-reflective feeling most evident in mood and emotion but always available 
	to consciousness; 	
	\item \textbf{Emotional episode} a complex set of interrelated sub-events concerned 
	with a specific object;
	\item \textbf{Mood} - the appropriate designation for affective states that are about 
	nothing specific or about everything - about the world in general.	
\end{itemize}
Obviously a key difference between Moods and Emotions is the fact that moods usually last 
longer. What matters from our perspective is to use well-defined models which represent the 
mood states of a person and the mood categories of songs. One of most used in literature is 
Russel's planar model \cite{russell1980circumplex} shown in Fig. 1.1.
This model is based on two dimensions: \emph{Valence} (pleasant-unpleasant) and 
\emph{Arousal} (aroused-sleepy). Valence represents how much an emotion is perceived 
as positive or negative whereas Arousal indicates how strongly the emotion is felt.
\begin{figure}
	\centering
	\begin{minipage}{.5\columnwidth}
		\centering
		\includegraphics[width=.95\columnwidth]{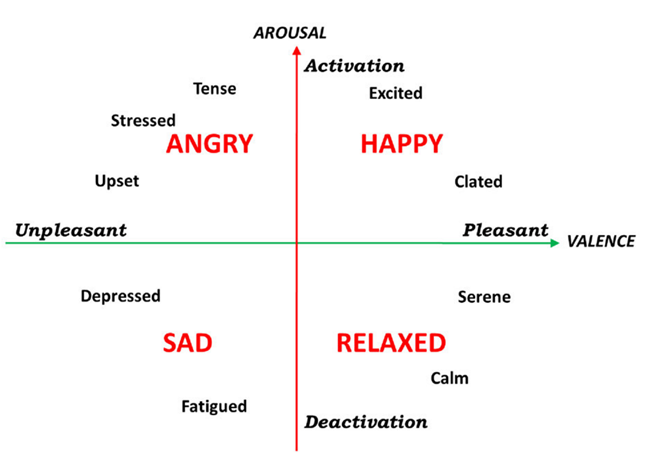}
		\caption{Circumplex model of mood}
		\label{fig:mockup1}
	\end{minipage}%
	\begin{minipage}{.5\columnwidth}
		\centering
		\includegraphics[width=.95\columnwidth]{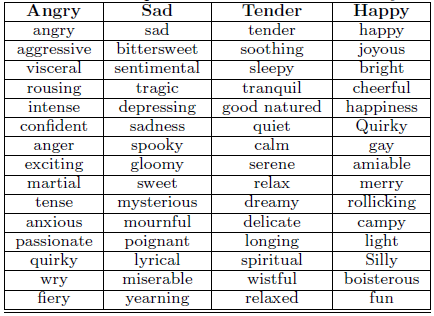}
		\caption{Representation of Folksonomy}
		\label{fig:mockup2}
	\end{minipage}
\end{figure}
A prominent categorical model for the mood of songs is \cite{Hevner1936}, which uses 66 descriptors 
categorized in 8 groups. A more recent approach we found is a folksonomy with four 
clusters of social community tags described in \cite{1466}. It is the one we use 
in this project.   
\subsection{Mood Recognition and Context} 
Psychologists evaluate the impact of music listening in one's mood using a standard 
method called Musical Mood Induction Process (MMIP). It consists of replaying 
mood-eliciting music to research participants. In \cite{vastfjall2002emotion} the 
authors outline several MMIP techniques for estimating the effects on mood given by music. 
Some of the most important ones are \emph{Self-reports}, \emph{Behavioral measures} and 
\emph{Physiological measures}. For this project we base the driver's mood recognition 
in physiological measures obtained from wearable sensors. The physiological parameters 
provided by wearables can be combined and used as valuable emotion-related data sources. 
For example, in \cite{Peter2005} the authors present one such system capable of 
recognizing emotions which is based on heart rate, skin conductivity, and skin 
temperature.
In this project we utilize heart rate dynamics to assess the arousal and valence the 
driver perceives when listening to the recommended songs. His/Her mood can be considered 
as a contextual parameter. In context-aware systems research, one of the definitions 
of context-aware computing is the ability of a mobile user's application to discover 
and react to changes in the environment (i.e. driver's mood state) they are situated 
in \cite{313011}. Context is defined by different concepts or factors such as time 
and/or date, weather conditions and  user's emotional state.   
\par
One approach used to consider the contextual factors in providing recommendations is 
the multidimensional model which considers dimensions as Cartesian products of some 
attributes and the recommendation space as Cartesian product of the dimensions \cite{Adomavicius:2005:ICI:1055709.1055714}. 
In a music recommender based on mood the dimensions could be 
\textit{\textbf{User} $\subseteq$ Uname x Age x Proffession},
\textit{\textbf{Item} $\subseteq$ SongTitle x Artist x Genre x MoodLabel}, 
\textit{\textbf{Context} $\subseteq$ Time x Location},
\textit{\textbf{Mood} $\subseteq$ DrivingStyle x HeartRate x SkinConductance}.
The other side of the coin is to detect or recognize mood categories inside musical 
tracks. In the literature this problem is addressed using the following approaches:
\begin{itemize}
	\item \textbf{Social Tags} - many tags (such as \emph{passionate}, \emph{autumnal}, \emph{witty}, 
	\emph{cool}) may be useful for mood or genre recognition in songs;
	\item \textbf{Lyrics} - song lyrics can also be used to classify music tracks into 
	mood categories \cite{Hu09lyric-basedsong, journals/ijcpol/XiaWW08};
	\item \textbf{Audio} - audio processing techniques were the earliest employed to 
	recognize mood in music using features like timbre, harmony, register, rhythm; 
	\item \textbf{Multimodal} - to attain better predictive accuracy many studies combine the 
	above three approaches in different ways building multi-modal algorithms 
	\cite{hu2010improving, Yang2008}.  
\end{itemize}
There are also several studies which try to shed light on the moods that different  
structural properties of music induce to people. According to \cite{10.3389/fpsyg.2013.00487} 
\emph{mode} is the highest important music property followed by \emph{tempo}, \emph{register}, 
\emph{dynamics}, \emph{articulation} and \emph{timbre}. All these properties of music exhibit 
certain correlations with mood categories. 
\subsection{Mood and Driving Style}
Several studies show correlations between a driver's mood and his/her behavior 
while driving: for instance, \cite{garrity} shows the influence on 
driving cautiousness given by moods like anger or depression; \cite{VanDerZwaag} shows the 
outcomes of an experiment conducted on drivers in which 
the sad and relaxed moods were found as connected to a safer driving; conversely, 
\cite{Nesbit} highlights the correlation between an angry mood and an aggressive style 
of driving. Regulative efforts should take place when the current driver's mood is not safe 
for driving: \cite{VanDerZwaag} shows the effectiveness of changing the mood of the 
listened songs in doing so and underlines that gradual shifts can obtain such result 
more efficiently.
In general, the actions that can be performed on the driver's mood through music 
recommendations are:
\begin{itemize}
\item \textbf{Mood Regulation} - a target desired mood, different from the current one, 
is set as goal;
\item \textbf{Mood Maintenance} - the driver is already in a suitable mood, thus actions 
have to be taken to try to maintain it.
\end{itemize}
\par
Driving style can be retrieved dynamically gathering telemetrics from the car. 
To do so, the OBD-II technology can be leveraged, %
since it provides access to a set of diagnostic  
information about the car functioning. %
Recently, a vast assortment of OBD-II adapters (the so-called \emph{dongles}) has been 
made available in the market. They may provide APIs to mobile applications, so they can 
telemetrics service and keep track of the driver's behavior.
An alternative to the use of OBD-II adaptors, as discussed in \cite{meng}, is the use of 
approximations done by the smarphones themselves, leveraging data collected from GPS, 
accelerometers and gyroscopes. Such measures, however, are vulnerable to external interference 
and provide little accuracy in short time windows. 
Once telemetric data are extracted and used to characterize the driver's behavior, they can 
be used to check whether the playlists recommended by the system have accomplished the 
objective of making the driving style shift to a safer one.   
\subsection{On-car Infotainment technology}
Lately, a rapid growth and diffusion of on-car infotainment platforms \cite{viereckl} 
has been witnessed. Car dashboards are now being connected to the Internet 
and/or integrated with driver's hand held devices, thus enabling finer music listening 
experiences than just the plain classic radio playback. What follows is a quick overview 
on some available solutions for the access to music streaming on car:
\begin{itemize}

\item \emph{Direct use of the smartphone} - The simplest solution is to connect the 
smartphone to the car speakers (via AUX or Bluetooth cables), and make use of it 
to provide the only human-machine interface with which the users will interact. ``Auto'' versions 
of renowned streaming music players (e.g., Spotify or Pandora), car-oriented music players 
accessing playlists stored in the smartphone, as well as prototypes of music recommendation 
apps \cite{Baltrunas2011} are available in the market.
\item \emph{Android Auto Integration} - Once a device is connected via a USB cable or 
bluetooth to a compatible car dashboard, it allows the use of applications installed on it 
as providers of music playlists to be played from the car speakers. The connectivity of the 
smartphone can be used to stream playlists. Once connected, the smartphone screen goes 
black and user interactions are performed with the car dashboard only. Android Auto 
mandates a common very minimal user interface for all the applications working with it.
\item \emph{Proprietary applications installed on dashboards} - Some automakers have 
signed partnerships with music streaming services provider, thus native streaming 
applications are installed on their dashboards. 
\item \emph{Mirrorlink} - It performs a mirroring of compatible applications installed on 
the smartphone. All human-machine interaction is performed through the car dashboard. 
The guidelines for the application graphics are really strict, and both the smartphone 
and the car must be MirrorLink-enabled.
\end{itemize}

%% file: sections/SystemDescription.tex
\section{System Description}
The architecture of the system is presented in fig. 1.3. 
\begin{figure}[!t]
	\centering
	\includegraphics[width=\columnwidth]{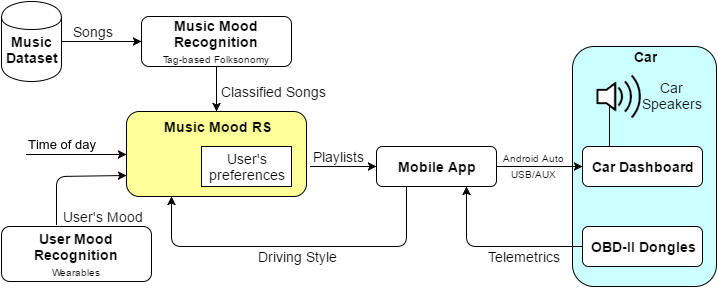}
	\caption{Schematic view of the system.}
	\label{fig:systemscheme}
\end{figure} 
The main module is the Music Mood RS, which is responsible for providing the 
appropriate playlist of songs recommended for the driver. This module 
analyzes many data such as driving style patterns, contextual factors, mood 
labeled musical tracks and driver's mood state. 
User Mood Detection is the module responsible for providing the driver's mood 
representation. It takes in mood related physiological 
data provided by wearable sensors (e.g.,  heart rate dynamics), computes 
the \emph{Valence} and \emph{Arousal} values and feeds the driver's mood 
state in the RS module.   
The OBD-II module is responsible for providing car telemetry data from which
driving style patterns are extracted. 
Music Mood Recognition module adds mood labels to the musical track the 
system uses.
The Mobile App module is an Android application which shows up the 
user interface and enables media playback. The application may be used 
on the smartphone itself or may be integrated with the car-dashboard after 
having connected the smartphone to it.
\subsection{Music Mood Recognition}
This module works on the many songs obtained from public music datasets such as 
last.fm, Yahoo Music Ratings, or Million Song Dataset. These datasets are very 
important for training and testing purposes and are further described in
\cite{7325106}. We use a classifier that is based on a semantic mood 
model described in \cite{1466}. The model was derived from social tags of songs 
collected from last.fm music community. 
It is in consonance with many other existing mood models. The 
four categories \emph{angry}, \emph{sad}, \emph{tender} and \emph{happy} can be seen as the representative 
mood categories of each of the four subplanes in the planar model of Russell. Other 
studies such as \cite{DBLP:conf/icmcs/SaariBFES13} also confirm that tag based semantic 
mood models are effective to predict perceived mood of the songs. The top tags of 
each category are presented in fig. 1.2.
\subsection{User Mood Detection}
We use the system presented in \cite{valenza2014revealing} to recognize the mood 
state of the driver. The authors use cardiovascular dynamics (Heart Rate Variability)
observations on short-time emotional stimuli. They utilize the images of International 
affective picture system (IAPS) described in \cite{citeulike:10370317} to provoke 
emotional stimulus and observe the physiological consequences. 
The emotional model used in their work is the Circumplex Model of Affect (CMA) 
which is basically the same model we adopt. Using only heartbeat 
dynamics they effectively distinguish between the two basic levels of both 
arousal and valence, thus allowing for the assessment of four basic emotions. 
An important advantage of their framework that has relevance for us is the fact 
that it is fully personalized and does not require data from a representative 
population of subjects. 
\subsection{Driving Style Recognition}
To track user's driving style, we have chosen to rely on the OBD-II 
adapter approach. In particular, we do a simple estimation of 
the driver's aggression in his/her driving style by calculating the jerkiness of the 
speed and acceleration profile of the car. To obtain the necessary telemetry data we leverage
the \emph{stats} API provided by Dash: the call provides the average speed in a 
finite time interval so the acceleration can be computed based on two subsequent 
measurements.
A heuristic threshold is utilized to discriminate between calm and aggressive driving 
style, based on estimations of the jerk (the first-order derivative of 
acceleration). We consider the thresholds provided by \cite{Murphey09}, obtained as 
the average jerk values on a number of drive cycles in typical scenarios.
The driving style is tagged \emph{aggressive} if the actual jerk is greater than the threshold 
correspondent to the scenario where the drive is having place.
The derived driving style is computed by the mobile application that is running the player 
and is then used as a flag inside the recommender to confirm or dissent   
the mood estimation of the wearables endorsed by the user (since an 
aggressive driving style is related to a high level of arousal, and in particular to an 
angry mood). Moreover, the driving style itself is used as a direct proof of the 
effectiveness of the recommendations, as they are used with the aim of relieving the 
driver from stress and aggression.
\subsection{Music Mood RS}
The recommendation module collects the different contextual,
user and song data, analyzes them and generates the most appropriate playlist.  
Time of day is a contextual parameter related to the rate of arousal the driver needs.
We assume that during the day there is no need for extra stimulation of the driver. 
For this reason the default recommended mood category is \emph{tender} which based on the driving
literature background (2.3) is the most favorable to a comfortable and safe driving. 
On the contrary, during a night drive it is usually better to avoid a sleepy state 
of mood and keep the driver more alerted, thus recommending \emph{happy} music. 
\par
We use the two values (arousal and valence) provided by the wearable sensors to identify 
the mood category the driver is in. The objective is to keep the driver in a relaxed state 
of mood by recommending music of his/her tastes. When the user is already in relaxed mood 
the recommender gives more priority to his/her past musical preferences. The driving style 
is obtained from the telemetric data of the OBD-II module and provides insights about how 
safe the car being driven. It falls in two categories, \emph{aggressive} and \emph{non aggressive}. 
The last data that goes in the recommendation module are the musical tracks which have 
already been classified in the four mood categories by the tag-based classifier. The 
recommender considers the contextual data and the  user's past preferences to generate 
the playlist. The driver is free to chose which song to play. If no selection is made 
the top ranked song is played automatically. 
\subsection{Mobile App}
\begin{figure}[!t]
\centering
\begin{minipage}{.5\columnwidth}
  \centering
  \includegraphics[width=.95\columnwidth]{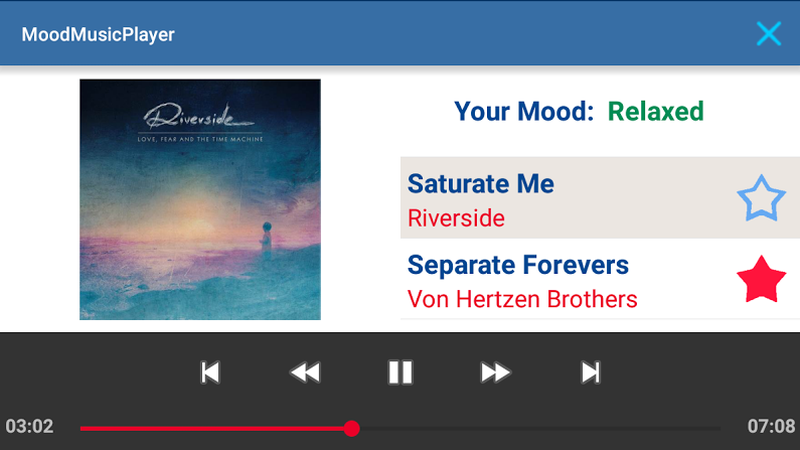}
	\caption{Mobile Application interface.}
	\label{fig:mockup1}
\end{minipage}%
\begin{minipage}{.5\columnwidth}
  \centering
  \includegraphics[width=.95\columnwidth]{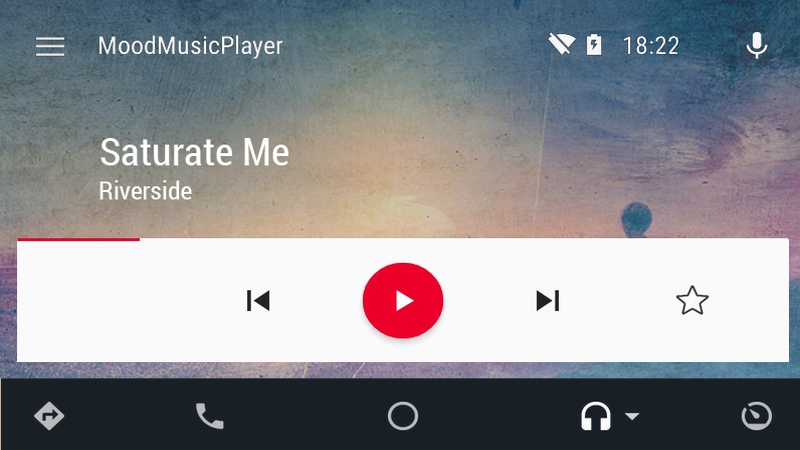}
	\caption{Android Auto interface.}
	\label{fig:mockup2}
\end{minipage}
\end{figure}

The main outcome of the project is an Android application providing a music player 
Service. Since the intended use of the app is inside an in-car environment, we made 
the application compatible with the Android Auto platform. However, full functionalities 
are also available if the car dashboard is not compatible to Android Auto. The phone 
should anyhow be connected via AUX/USB-cable to the car dashboard to enable music 
playback through the car speakers
The mobile app is targeted to Android 5.0 Lollipop and presents a simple landscape 
interface through which the user can give traditional music playback inputs and express 
his/her approval about the recommended songs. The application also shows information 
about the user's estimated mood. The user interface (see fig. \ref{fig:mockup1}) 
is designed to provide minimum distraction to the driver. 

The app is compatible with the Android Auto platform for the in-car streaming. To the 
standard interface, a button allowing to express appreciation for a recommended song 
has been added. The interface of the application can be seen in fig. \ref{fig:mockup2}.

%% file: lnicst.bbl
\begin{thebibliography}{10}

\bibitem{Adomavicius:2005:ICI:1055709.1055714}
G.~Adomavicius, R.~Sankaranarayanan, S.~Sen, and A.~Tuzhilin.
\newblock Incorporating contextual information in recommender systems using a
  multidimensional approach.
\newblock {\em ACM Trans. Inf. Syst.}, 23(1):103--145, Jan. 2005.

\bibitem{Baltrunas2011}
L.~Baltrunas, M.~Kaminskas, B.~Ludwig, O.~Moling, F.~Ricci, A.~Aydin, K.-H.
  L{\"u}ke, and R.~Schwaiger.
\newblock {\em E-Commerce and Web Technologies: 12th International Conference,
  EC-Web 2011, Toulouse, France, August 30 - September 1, 2011. Proceedings},
  chapter InCarMusic: Context-Aware Music Recommendations in a Car, pages
  89--100.
\newblock Springer Berlin Heidelberg, Berlin, Heidelberg, 2011.

\bibitem{7325106}
E.~\c{C}ano and M.~Morisio.
\newblock Characterization of public datasets for recommender systems.
\newblock In {\em Research and Technologies for Society and Industry Leveraging
  a better tomorrow (RTSI), 2015 IEEE 1st International Forum on}, pages
  249--257, Sept 2015.

\bibitem{dibben2007exploratory}
N.~Dibben and V.~Williamson.
\newblock {An exploratory survey of in-vehicle music listening}.
\newblock {\em Psychology of Music}, 35(4):571, 2007.

\bibitem{10.3389/fpsyg.2013.00487}
T.~Eerola, A.~Friberg, and R.~Bresin.
\newblock Emotional expression in music: Contribution, linearity, and
  additivity of primary musical cues.
\newblock {\em Frontiers in Psychology}, 4(487), 2013.

\bibitem{tenenbaummeasurement}
P.~Ekkekakis.
\newblock {\em Measurement in Sport and Exercise Psychology}, chapter Affect,
  Mood, and Emotion.
\newblock Human Kinetics, 2012.

\bibitem{garrity}
R.~D. Garrity and J.~Demick.
\newblock Relations among personality traits, mood states, and driving
  behaviors.
\newblock {\em Journal of Adult Development}, 8(2):109--118, April 2001.

\bibitem{Hevner1936}
K.~Hevner.
\newblock {Experimental studies of the elements of expression in music}.
\newblock {\em The American Journal of Psychology}, 48:246--268, 1936.

\bibitem{hu2010improving}
X.~Hu.
\newblock {\em Improving music mood classification using lyrics, audio and
  social tags}.
\newblock PhD thesis, Citeseer, 2010.

\bibitem{Hu09lyric-basedsong}
Y.~Hu, X.~Chen, and D.~Yang.
\newblock Lyric-based song emotion detection with affective lexicon and fuzzy
  clustering method.
\newblock In {\em Proceedings of ISMIR 2009}, pages 123--128, 2009.

\bibitem{citeulike:10370317}
P.~J. Lang, M.~M. Bradley, and B.~N. Cuthbert.
\newblock {International affective picture system (IAPS): Affective ratings of
  pictures and instruction manual}.
\newblock Technical Report A-8, The Center for Research in Psychophysiology,
  University of Florida, Gainesville, FL, 2008.

\bibitem{1466}
C.~Laurier, M.~Sordo, J.~Serr{\`a}, and P.~Herrera.
\newblock Music mood representations from social tags.
\newblock In {\em International Society for Music Information Retrieval (ISMIR)
  Conference}, pages 381--386, Kobe, Japan, 26/10/2009 2009.

\bibitem{meng}
R.~Meng, C.~Mao, and {Romit Roy Choudhury}.
\newblock Driving analytics: Will it be obds or smartphones?
\newblock {\em Zendrive Whitepaper}, 2014.

\bibitem{Murphey09}
Y.~L. Murphey, R.~Milton, and L.~Kiliaris.
\newblock Driver's style classification using jerk analysis.
\newblock In {\em Computational Intelligence in Vehicles and Vehicular Systems,
  2009. CIVVS '09. IEEE Workshop on}, pages 23--28, March 2009.

\bibitem{Nesbit}
S.~M. Nesbit, J.~C. Conger, and A.~J. Conger.
\newblock A quantitative review of the relationship between anger and
  aggressive driving.
\newblock {\em Aggression and Violent Behavior}, 12(2):156 -- 176, 2007.

\bibitem{Peter2005}
C.~Peter, E.~Ebert, and H.~Beikirch.
\newblock {\em Affective Computing and Intelligent Interaction: First
  International Conference, ACII 2005, Beijing, China, October 22-24, 2005.
  Proceedings}, chapter A Wearable Multi-sensor System for Mobile Acquisition
  of Emotion-Related Physiological Data, pages 691--698.
\newblock Springer Berlin Heidelberg, Berlin, Heidelberg, 2005.

\bibitem{russell1980circumplex}
J.~Russell.
\newblock {A circumplex model of affect}.
\newblock {\em Journal of personality and social psychology}, 39(6):1161--1178,
  1980.

\bibitem{DBLP:conf/icmcs/SaariBFES13}
P.~Saari, M.~Barthet, G.~Fazekas, T.~Eerola, and M.~B. Sandler.
\newblock Semantic models of musical mood: Comparison between crowd-sourced and
  curated editorial tags.
\newblock In {\em 2013 {IEEE} International Conference on Multimedia and Expo
  Workshops, San Jose, CA, USA, July 15-19, 2013}, pages 1--6, 2013.

\bibitem{schafer2013psychological}
T.~Sch{\"a}fer, P.~Sedlmeier, C.~St{\"a}dtler, and D.~Huron.
\newblock The psychological functions of music listening.
\newblock {\em Frontiers in psychology}, 4(511):1--33, 2013.

\bibitem{313011}
B.~N. Schilit and M.~M. Theimer.
\newblock Disseminating active map information to mobile hosts.
\newblock {\em IEEE Network}, 8(5):22--32, Sept 1994.

\bibitem{valenza2014revealing}
G.~Valenza, L.~Citi, A.~Lanat{\'a}, E.~P. Scilingo, and R.~Barbieri.
\newblock Revealing real-time emotional responses: a personalized assessment
  based on heartbeat dynamics.
\newblock {\em Scientific reports}, 4, 2014.

\bibitem{VanDerZwaag}
M.~D. van~der Zwaag, J.~H. Janssen, C.~Nass, J.~H. Westerink, S.~Chowdhury, and
  D.~de~Waard.
\newblock Using music to change mood while driving.
\newblock {\em Ergonomics}, 56(10):1504--1514, 2013.
\newblock PMID: 23998711.

\bibitem{vastfjall2002emotion}
D.~V{\"a}stfj{\"a}ll.
\newblock Emotion induction through music: A review of the musical mood
  induction procedure.
\newblock {\em Musicae Scientiae}, 5(1 suppl):173--211, 2002.

\bibitem{viereckl}
R.~Viereckl, D.~Ahlemann, A.~Koster, and S.~Jursch.
\newblock Racing ahead with autonomous cars and digital innovation.
\newblock {\em Auto Tech Review}, 4(12):18--23, dec 2015.

\bibitem{journals/ijcpol/XiaWW08}
Y.~Xia, L.~Wang, and K.-F. Wong.
\newblock Sentiment vector space model for lyric-based song sentiment
  classification.
\newblock {\em Int. J. Comput. Proc. Oriental Lang.}, 21(4):309--330, 2008.

\bibitem{Yang2008}
Y.-H. Yang, Y.-C. Lin, H.-T. Cheng, I.-B. Liao, Y.-C. Ho, and H.~H. Chen.
\newblock {\em Advances in Multimedia Information Processing - PCM 2008: 9th
  Pacific Rim Conference on Multimedia, Tainan, Taiwan, December 9-13, 2008.
  Proceedings}, chapter Toward Multi-modal Music Emotion Classification, pages
  70--79.
\newblock Springer Berlin Heidelberg, Berlin, Heidelberg, 2008.

\end{thebibliography}
